\documentclass[amsmath,amssymb,aps,twocolumn]{revtex4-1}

\usepackage[utf8]{inputenc}
\usepackage{graphicx}
\usepackage{dcolumn}
\usepackage{bm}

\newcommand{\cs}{\langle\sigma\upsilon\rangle}

\begin{document}

\title{Coy Dark Matter and the anomalous magnetic moment}

\author{Andi Hektor}
 \affiliation{National Institute of Chemical Physics and Biophysics, Akadeemia tee 23, 12618 Tallinn, Estonia}
\author{Luca Marzola}
 \affiliation{Laboratory of Theoretical Physics, Institute of Physics, University of Tartu; Tähe 4, 51010 Tartu, Estonia}

\begin{abstract}
Coy Dark Matter removes the tension between the traditional WIMP paradigm of Dark Matter and the latest exclusion bounds from direct detection experiments. In this paper we present a leptophilic Coy Dark Matter model that, on top of explaining the spatially extended 1-5~GeV $\gamma$-ray excess detected at the Galactic Center, reconciles the measured anomalous magnetic moment of muon with the corresponding Standard Model prediction. The annihilation channel of DM is $\chi\chi \to \tau\bar\tau$ with the DM mass $m_\chi = 9.43\,(^{+.063}_{-0.52}\,{\rm stat.})\,(\pm 1.2 \,{\rm sys.})$~GeV given by best-fit of the $\gamma$-ray excess. Fitting the measured anomalous magnetic moment of the muon requires instead a pseudoscalar mediator with a minimal mass $m_a = 12^{+7}_{-3}$~GeV.
\end{abstract}

\maketitle

\section{Introduction}

According to the present understanding, the matter content of the Universe is dominated by a very weakly interacting component: the Dark Matter (DM). In the most promising scenarios, DM consists of a thermal relic density of stable and weakly interacting massive particles (WIMPs). In fact, miraculously, particles with masses and annihilation cross sections set by the electroweak scale automatically yield the observed value for DM density, through the freeze-out mechanism (for a review, see~\cite{Jungman:1995df,Bertone:2004pz}). On the experimental side, the WIMP paradigm motivates the efforts aimed to the detection of three types of DM signature: elastic scattering between DM and SM particles, production of DM particles at colliders and annihilation of DM particles in the Universe. In particular, for the latter case, it is essential to search for direct annihilation signals from dense DM regions, as well as  studying the implied indirect large scale effects that, for example, would affect CMB.

In this regard, the center of our Galactic DM halo at the Galactic Center (GC) should provide the strongest annihilation signal. Unfortunately, GC also harbours an extremely dense environment filled with stars, stellar relics and related cosmic rays, dust and gas. As a consequence the possible annihilation signals can easily be disguised as result of active astrophysical processes. In spite of that, different hints of DM annihilation have been reported in literature in recent years pointing towards the DM annihilation cross sections at the order of the thermal freeze-out one, $\cs_{\rm th}$. In 2008, the PAMELA satellite mission measured an excess of cosmic positrons above the energy of 20~GeV, later confirmed by Fermi LAT and AMS-02 \cite{Adriani:2008zr,Abdo:2009zk,Aguilar:2013qda}. In 2009, instead, the public data of Fermi LAT~\cite{Atwood:2009ez} showed a spatially extended $\gamma$-ray excess at 1-5~GeV at GC~\cite{Goodenough:2009gk,Hooper:2010mq,Abazajian:2010zy,Boyarsky:2010dr,Hooper:2011ti,Abazajian:
2012pn,Gordon:2013vta,Macias:2013vya,Abazajian:2014fta,Daylan:2014rsa,Lacroix:2014eea}. In 2012, a hint of a $\gamma$-ray line(s) at 130~GeV was found \cite{Bringmann:2012vr,Weniger:2012tx,Tempel:2012ey,Su:2012ft}.

On the other hand, strong constraints from XENON100 [8] and LUX [9] direct detection experiments recently excluded scattering cross sections near the typical weak-scale value. Hence, with the new exclusion bounds approaching the predicted region, both the above claims and the WIMP paradigm itself have started to tremble. Fortunately, it is still possible to construct rather natural particle physical models that possess an annihilation cross section large enough to explain the detected signal, but, at the same time, present a suppressed DM-nucleon scattering cross section. A minimal and elegant example is provided by the so-called ``Coy Dark Matter'' (CDM), recently proposed by Boehm~et~al~\cite{Boehm:2014hva}. In this model, the DM particle is a Dirac fermion which interacts with the Standard Model (SM) particles by the exchange of a relatively light pseudoscalar mediator. The new couplings to the SM particles are assumed to be proportional to the corresponding Higgs Yukawa couplings, as motivated by minimal 
flavor violation [42]. Hence, if the mass of the DM particle $\chi$ is below the mass of top quark, the dominant annihilation channel is $\chi\chi \to \bar{b}b$. With the choice $m_\chi \simeq 30$~GeV the spatially extended $\gamma$-ray excess at 1-5~GeV at GC can be fitted for a natural value of the DM annihilation cross section $\sim\cs_{\rm th}$~\cite{Gordon:2013vta,Macias:2013vya,Abazajian:2014fta,Daylan:2014rsa}. Alternatively, the $\chi\chi \to \bar{\tau}\tau$ channel can fit the signal, provided a lower mass for $\chi$ is adopted: $m_\chi \simeq 10$~GeV. In this case the SM Yukawa structure adopted for the SM-pseudoscalar couplings in the CDM model is also to be modified. For instance, a viable model is achieved by assuming leptophilic SM-pseudoscalar couplings, i.e. by maintaining the Yukawa structure only for the SM leptons and by neglecting the coupling of the pseudoscalar mediator to quarks.

Intriguingly, a light pseudoscalar coupled to muons, as proposed by CDM, results in a new contribution to the anomalous magnetic moment of this particle, $a_\mu$~\cite{Chang:2000ii}. To this regard, among DM, neutrino masses and baryon asymmetry, the $\sim 3.4\sigma$ deviation of $a_\mu$ from the SM value is a compelling experimental evidence that points to physics beyond SM (for a review see~\cite{Beringer:1900zz} and references therein). In this study, we calculate the anomalous magnetic moment of electron, muon and tau in the framework of CDM, aiming to constrain and fit both the measured value of $a_\mu^{\rm obs}$ and the $\gamma$-ray excess at 1-5~GeV from GC. We will show that the pseudoscalar-muon coupling has to be enhanced by a factor ${\cal O}(100)$ to fit $a_\mu^{\rm obs}$. Within the CDM framework this if forbidden by the studies on the $\Upsilon$ resonance decays, which constrain the pseudoscalar-muon coupling below the required value \cite{Aaltonen:2009rq,Aubert:2009cka,Schmidt-Hoberg:2013hba}. 
However, this constraint is trivially avoided in case of leptophilic CDM, where the $\chi\chi \to \bar{\tau}\tau$ channel is responsible for fitting the 1-5~GeV $\gamma$-ray excess at GC. We will therefore show that the leptophilic CDM scenario allows to explain $a_\mu^{\rm obs}$, to fit the $\gamma$-ray excess at 1-5~GeV from GC and to avoid the remaining well known constraints.

The paper is structured as follows: in Sec. II we calculate the anomalous magnetic moment of muon, electron and tau within the CDM framework. In Sec. III discuss the constraints on CDM from $a_{e,\mu,\tau}$ and the resulting scenario. Finally, in Sec. IV we draw our conclusions.

\section{Anomalous magnetic moment induced by a pseudoscalar}

The prototype model we consider in this study is the CDM one, presented by Boehm~et~al~\cite{Boehm:2014hva}. DM is composed by fermions $\chi$ with mass $m_{\chi}$ having interaction with the SM content via a pseudo-scalar field $a$ of mass $m_a$. The interaction is governed by the Lagrangian
\begin{equation}\label{eq:lag}
{\cal L} \supset - i\frac{g_{\chi}}{\sqrt 2} \, a \bar\chi \gamma^5 \chi - i\sum_{f} \frac{g_f}{\sqrt 2} \, a \bar f \gamma^5 f + \text{ h.c.}
\end{equation} 
where $f$ runs on the SM fermions and 
\begin{equation}\label{eq:defA}
g_f \equiv A \, y_f = A\,\frac{\sqrt 2 \, m_f}{v} 
\end{equation}
being $v = 246$~GeV, the Higgs boson vacuum expectation value. The factor $A$ is set to 1 in the original approach by Boehm~et~al~\cite{Boehm:2014hva}. Naturally, for $A \equiv 1$, the couplings $g_f$ replicate the exact structure of the SM Yukawa couplings. However, as we will now show, in order to fit $a_\mu^{\rm obs}$ it is necessary to impose $A > 1$, amplifying by net the SM Yukawa hierarchy.

The anomalous magnetic moment of a fermion can be quantified in
\begin{equation}
a_f \equiv \frac{g_f - 2}{2}.
\end{equation}
Clearly, for the Lagrangian~\eqref{eq:lag}, at one loop level the magnetic moment of all the fermions receive a new contribution from the exchange of a pseudo-scalar particle, which amounts to \cite{Chang:2000ii}:
\begin{align}\label{eq:delta1loop}
\delta a_f^{(1)} &= -\frac{m_f^2}{8\pi^2 m_a^2} \left(\frac{m_f}{v}\right)^2 \, H\left(\frac{m_f^2}{m_a^2}\right),\\
H(x) &\equiv \int\limits_0^1 \frac{y^3}{1 - y + y^2 x}\, \, dy.
\end{align}
Notice that 
\begin{align}
H(x) &\xrightarrow{0<x\ll1} -\ln(x)-\frac{11}{6}>0,\\
H(x) &\xrightarrow{x\to\infty} 0
\end{align} 
therefore $\delta a_f^{(1)} < 0$.
At the two loop level, the pseudo-scalar mediator brings two further contributions to the anomalous coupling of the fermions (see Fig. 1 in~\cite{Chang:2000ii}). These can be quantified in
\begin{align}\label{eq:2loopg}
\delta a_f^{(2)\gamma} &= \frac{\alpha^2}{8\pi^2 \sin^2\theta_W} \frac{m_f^2}{m^2_W}\sum_{x=t, b, \tau} N_c^x q_x^2 \frac{m_x^2}{m_a^2} \mathcal{F}\left(\frac{m^2_x}{m^2_a}\right),\\ 
\mathcal{F}(x) &\equiv \int\limits_0^1 \frac{\ln\left(\frac{x}{z(1-z)}\right)}{x-z(1-z)}\, \, dz
\end{align}
and
\begin{equation} \label{eq:2loopz}
\begin{split}
\delta a_f^{(2)Z} &= \frac{\alpha^2 g_V^f}{8\pi^2 \sin^4\theta_W \cos^4\theta_W}\frac{m_f^2}{m_Z^2} \\
\quad &\cdot \sum_{x=t, b, \tau} N_c^x  q_x  g_V^x \frac{m_x^2}{m_Z^2 - m_a^2} \left[ \mathcal{F}\left(\frac{m^2_x}{m^2_Z}\right)-\mathcal{F}\left(\frac{m^2_x}{m^2_a}\right) \right],
\end{split}
\end{equation}
where $\alpha$ is the fine structure constant, $N_c^x$ the number of colours of the particle $x$ and $g_V^x$ its vector coupling:
\begin{equation}
g_V^x \equiv \frac{1}{2}I_3^x - q_x \sin^2\theta_W.
\end{equation}
The function $\mathcal{F}(x)$ is defined as
\begin{align}
\mathcal{F}(x) &\xrightarrow{0<x\ll1}\ln^2(x)+\frac{\pi^2}{3},\\
\mathcal{F}(x) &\xrightarrow{x\to\infty}\frac{\ln(x)+2}{x}.
\end{align} 

\begin{figure}[h]
\includegraphics[width=0.48\textwidth]{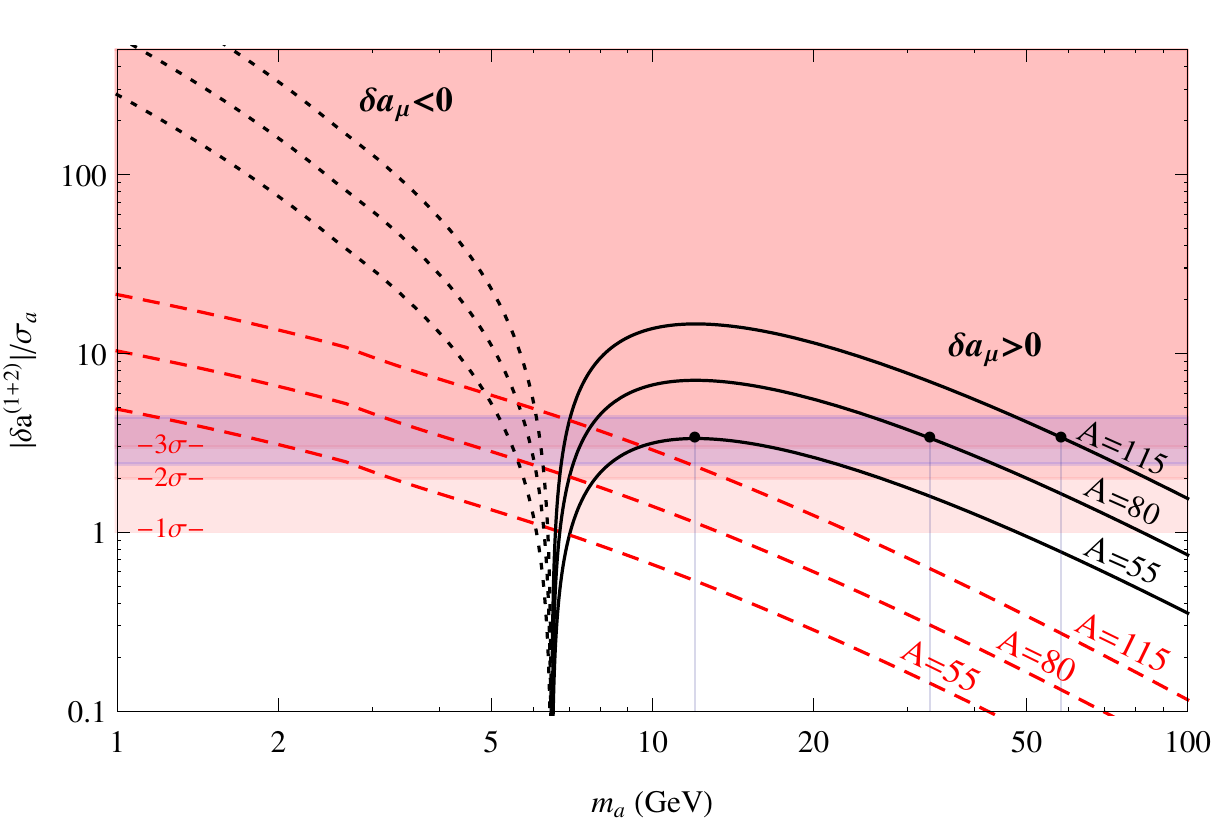}
\caption{The one and two-loop contributions of the pseudoscalar $a$ to the g-2 of muon (black) and electron (dashed red) given in units of the corresponding standard deviation. The contribution to the muon g-2 is negative for $m_a \lesssim 6.7$~GeV (dotted black). The pink regions quantify the deviation from the SM value of g-2. The blue region denotes the measured value of the muon g-2 (1$\sigma$). The black dots on the muon curves indicate the best fits of the experimental value of g-2, the 3.4$\sigma$ excess. The factor $A$ is defined in Eq. \ref{eq:defA}.}\label{fig1}
\end{figure}

Notice that when $A>1$ the above loop contributions are enhanced by a factor $A^2$. Fig. \ref{fig1} shows the absolute value of these additional contributions at one plus two loop level, $\delta a = \delta a_f^{(1)} + \delta a_f^{(2)\gamma} + \delta a_f^{(2)Z} \propto (A y_f)^2$, in units of the standard deviation of the corresponding anomalous magnetic moment. We present both the cases of electron and muon, while the tau is excluded due to the huge uncertainties that plague the measurement. The implication of our fitting and the possible constraints are discussed in the next section.

\section{The muon g-2, direct and indirect signatures in the context of Coy Dark Matter}

As made clear by Fig. \ref{fig1}, fitting the muon g-2 data requires a value of $A \gtrsim 50$. The minimum value of $A$ is $50$ for $m_a \simeq 12$~GeV. Above this threshold the required muon g-2 can be obtained for $m_a$ in two different ranges. For instance, if $A=80$ then the best fit masses are approximately $7$ and $32$~GeV. However, in all these cases, the lowest mass resulting form the fitting is always constrained by the electron g-2 data. Returning to our previous example, if $A=80$ then the electron g-2 deviates more than 3$\sigma$ from its measured values if $m_a \simeq 7$~GeV. The upper limit of $A$ comes from the unitary limit of $g_\tau \equiv A y_\tau \leq 4 \pi$, which gives $A_{\max} = 4 \pi v /(\sqrt{2} m_\tau) \simeq 1000$.

As anticipated, the searches for the direct production of the pseudoscalar mediator in $\Upsilon$ resonance decays, followed by the muonic decay of the former, constrain the pseudoscalar-muon coupling to values much below the  enhanced Yukawa structure required for $y_f$s~\cite{Aaltonen:2009rq,Aubert:2009cka,Schmidt-Hoberg:2013hba}. For example these experiments impose $A=1$ for $m_a=7$~GeV and $A=0.01$ for $m_a=5$~GeV~\cite{Schmidt-Hoberg:2013hba}. It is consequently mandatory to modify the SM Yukawa structure of the new couplings in order to fit the muon g-2. In this regard, the most elegant modification is assuming leptophilic CDM, where $g_f\approx0$ for quarks while $g_f = A y_f$ for the SM leptons. Hence,  we recalculated the g-2 enhancement factors under the assumptions of our leptophilic scenario, confirming that the conclusions shown in Fig. \ref{fig1} still hold.

With the leptophilic scenario in hand, we re-evaluate the constraints from LHC~\cite{Boehm:2014hva} and LEP~\cite{Fox:2011fx}. The monojet constraints from LHC are trivially avoided in the leptophilic model, because the pseudoscalar has no tree level coupling to quarks. If we assume $A={\cal O}(100)$ and fix $g_\chi$ by the fitting of the GC annihilation signal, ${\cs}_{\tau\bar\tau}^{\rm GC} = (0.51 \pm 0.24) \times 10^{-26}$~cm$^3$~s$^{-1}$~\cite{Abazajian:2014fta}, the constraint from $e^-e^+ \to \chi\chi$ results $\Lambda \equiv m_a/\sqrt{A y_e g_\chi} = {\cal O}({\rm TeV})$. The value is well above the current exclusion limits, e.g., see Fig. 2 in Ref.~\cite{Fox:2011fx}.

Several papers based on the public Fermi LAT data claim that a spatially extended $\gamma$-ray excess at GC favours DM annihilation into two possible channels: $\chi\chi \to \tau\bar\tau$ and $\chi\chi \to b\bar b$~\cite{Goodenough:2009gk,Hooper:2010mq,Abazajian:2010zy,Boyarsky:2010dr,Hooper:2011ti,Abazajian:2012pn,Gordon:2013vta,Macias:2013vya,Abazajian:2014fta,Daylan:2014rsa,Lacroix:2014eea}. For leptophilic CDM the main annihilation channel is obviously $\chi\chi \to \tau\bar\tau$, furthermore the $\mu$ and $e$ channels can be safely neglected as $y_\mu/y_\tau \sim 0.1$ and $y_e/y_\tau \sim 10^{-3}$. According to Lacroix~et~al~\cite{Lacroix:2014eea} the tau channel gives even a better fit once the secondary production of $\gamma$-rays (by inverse Compton and bremsstrahlung) is taken into account. In this study we adopt the best-fit values of Ref.~\cite{Abazajian:2014fta}: ${\cs}_{\tau\bar\tau}^{\rm GC} = 2 \times (0.51 \pm 0.24) \times 10^{-26}$~cm$^3$~s$^{-1}$ and $m_\chi = 9.43\,(^{+.63}_{-0.52}\,{\rm 
stat.})\,(\pm 1.2 \,{\rm sys.})$~GeV (similar values are quoted in~\cite{Lacroix:2014eea}). The leading factor 2 of the cross-section comes from the Dirac fermion nature of $\chi$ compared to the Majorana one in Ref.~\cite{Abazajian:2014fta}.

\begin{figure}[h]
\includegraphics[width=0.48\textwidth]{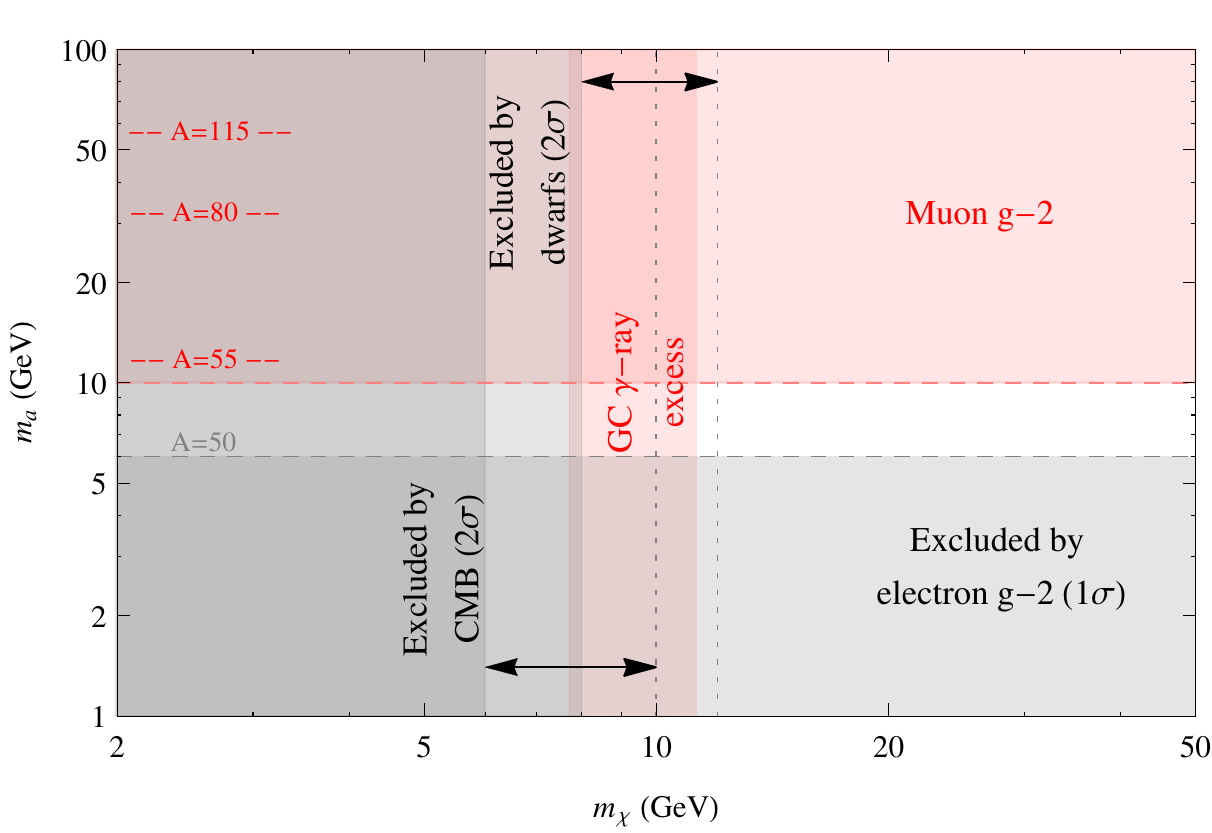}
\caption{The vertical pink stripe shows the best fit region (1$\sigma$) of the $\gamma$-ray excess via the $\tau$-channel at the Galactic Centre. The horizontal pink region denotes the best fit (1$\sigma$) for the muon g-2. The values of the $A$ parameter required by the fit for different masses $m_a$ are shown at the left axis. For the CDM model, the unitary limit, $A_{\max} \simeq 1000$, is very close to the top part of the frame. The gray regions and the dashed black lines show the constrains (both 2$\sigma$) from CMB~\cite{Hutsi:2010ai,Hutsi:2011vx,Madhavacheril:2013cna} and dwarf galaxies \cite{GeringerSameth:2011iw,Ackermann:2013yva} respectively.}\label{fig2}
\end{figure}

The best-fit mass $m_\chi$ is highlighted in Fig.~\ref{fig2} by the vertical pink band. Accordingly, the best-fit mass $m_a$ for the muon g-2 is shown by the horizontal pink region. The best-fit values of the factor $A$ are reported next to the ordinate axis. The vertical gray regions represent the constraints from  CMB~\cite{Hutsi:2010ai,Hutsi:2011vx,Madhavacheril:2013cna} and dwarf galaxies~\cite{GeringerSameth:2011iw,Ackermann:2013yva}, with the associated arrows denoting the variation of the latter throughout the references. There is clearly a tension between the constraints and the $\gamma$-ray signal from GC. However, we underline that the density profile of DM is (both theoretically and experimentally) poorly known at the very center of the Galaxy. Consequently, the present estimation of ${\cs}_{\tau\bar\tau}^{\rm GC}$ are hampered. We also remark that, in obtaining the results of Fig.~\ref{fig2}, we neglected the effects of the Sommerfeld enhancement for the mass 
region $m_a<m_\chi$  (which is constrained by the electron g-2 data in any case), as well as those due to a fine-tuned Breit–Wigner resonance enhancement at $m_a \approx 2 m_\chi$.

The constraints from direct detection experiments are very weak due to the leptophilic and CDM nature of our model~\cite{Boehm:2014hva}. In principle, the effects of electron-DM scattering could result in the signal observed by the DAMA/LIBRA experiment~\cite{Kopp:2009et}. However, unfortunately, in case of pseudo-scalar couplings the $e$-$\chi$ scattering cross section is suppressed by a factor $(m_e/m_\chi)^2 v^4 \approx 10^{-20}$, where $v\approx 10^{-3}$ (see Table 1 and Eq.~(10) in~\cite{Kopp:2009et}). Hence, even by imposing the maximum allowed value of $A = A_{\max} \simeq 1000$, the scattering rate remains below the required value. Solar constraints from DM capture and annihilation in the Sun are negligible for the same reason~\cite{Desai:2004pq,Wikstrom:2009kw,Kappl:2011kz}.

\section{Conclusions}

The CDM model removes the tension between the traditional WIMP paradigm of Dark Matter and the latest exclusion bounds from direct detection experiments~\cite{Boehm:2014hva}. We presented a leptophilic Coy Dark Matter model that, on top of explaining the spatially extended 1-5~GeV $\gamma$-ray excess detected at the Galactic Center, reconciles the measured anomalous magnetic moment of muon. Our results can be summarized as follows.
\begin{itemize}
 \item The $\gamma$-ray excess measured at GC is due to DM annihilation via the  $\chi\chi \to \tau\bar\tau$ channel, with the resulting best-fit mass $m_\chi \simeq 10$~GeV.
 \item Within our model, the measured value of the anomalous magnetic moment of the muon is due to the contribution of a light pseudoscalar mediator with mass $m_a = 12^{+7}_{-3}$~GeV. At this minimal mass, the pseudoscalar-muon coupling must be a factor $\sim 50$ larger than the muon Yukawa coupling.
 \item The leptophilic Coy Dark Matter scenario (which explains the measured anomalous magnetic moment of the muon) is compatible with the constraints from  CMB, dwarf galaxies, solar neutrinos, accelerator and direct detection experiments. The CMB and dwarf galaxy constraints are approaching the predicted best-fit value ${\cs}_{\tau\bar\tau}^{\rm GC}$ and might potentially rule out the model in near future.
\end{itemize}
~

\section{Acknowledgements}

This work was supported in part by grants IUT23-6, ETF8943 and by EU through the European Regional Development Fund and by ERDF project 3.2.0304.11-0313 Estonian Scientific Computing Infrastructure (ETAIS). LM acknowledges the European Social Fund for supporting his work under the grant MJD387. AH acknowledges the European Social Fund for the grants MTT8, MTT60 and the European Regional Development Fund for the grant TK120. The authors thank R.~Laha
for useful discussions.

\end{document}